\documentclass[aps,prl,showpacs,preprintnumbers,twocolumn,superscriptaddress]{revtex4-2}
\usepackage{amsmath,amssymb}
\usepackage{bm}
\usepackage{tipa}
\usepackage{upgreek}
\usepackage{comment}
\usepackage{mathrsfs}
\usepackage{graphicx}
\usepackage{braket}
\usepackage{enumitem}
\usepackage{mathbbol}
\usepackage{gensymb}
\usepackage[normalem]{ulem}
\usepackage{color}
\usepackage[colorlinks,bookmarks=true,citecolor=blue,linkcolor=red,urlcolor=blue]{hyperref}
\usepackage{hyperref}
\usepackage{dsfont}
\begin{document}

\title{Non-reciprocal Ising gauge theory}

\begin{abstract}
Non-reciprocity and geometric frustration enable many-body systems to avoid crystalline order and instead exhibit complex, liquid-like behavior.
Here we show that their interplay is richer than the sum of its parts, leading to surprising structural and dynamical phenomena. In our minimal model, two copies of Ising gauge theory are non-reciprocally coupled in a way that crucially preserves a local $\mathbb{Z}_2$ symmetry. 
We discover that the combined Wilson loop observable of the two copies exhibits linear asymptotic scaling, with a quasiparticle-pair confinement length tuned by the strength of the non-reciprocal coupling. 
Key dynamical features are revealed in the behavior of individual deconfined excitations due to strong interactions induced by the non-reciprocity, leading to motion on a critical percolation cluster that follows a self-avoiding trail. Mapping from this quasiparticle dynamics onto the magnetic noise spectrum, we discover that non-reciprocity tunes topological logarithmic contributions and causes long-lived metastable states due to quasiparticle trapping. 
Our work opens the way for broader investigations of geometrically frustrated non-reciprocity.
\end{abstract}

\author{Nilotpal Chakraborty}
\email{nc553@cam.ac.uk}
\affiliation{TCM Group, Cavendish Laboratory, University of Cambridge, Cambridge CB3 0HE, United Kingdom}

\author{Anton Souslov}
\affiliation{TCM Group, Cavendish Laboratory, University of Cambridge, Cambridge CB3 0HE, United Kingdom}

\author{Claudio Castelnovo}
\affiliation{TCM Group, Cavendish Laboratory, University of Cambridge, Cambridge CB3 0HE, United Kingdom}

\maketitle
%
%


Non-reciprocal interactions define a plethora of out-of-equilibrium systems, spanning statistical physics, neuroscience, social sciences, ecology, and open quantum systems~\cite{asymmneur,parisi1986asymmetric,derrida1987exactly,strog1,strog2,gamenonrec,bascompte2006asymmetric,clerk2022introduction}. 
Nevertheless, a framework for understanding their behavior in many-body systems is only starting to be developed~\cite{lorenzana2025nonreciprocity,fruchartreview}. 
Non-reciprocity was recently shown to give rise to long-lived oscillations between ordered states in spin models with local interactions~\cite{NRIsing,DynphaseNRIsing}. Microscopically, this non-reciprocal dynamics breaks detailed balance and lies outside the conventional treatment of phase-transitions and critical phenomena~\cite{fruchart2021non}. As a consequence, non-reciprocal many-body systems can exhibit a rich interplay of synchronization,  ordering~\cite{you2020nonreciprocity,Duannr,hongstrogprl,supriyonr,Ivlevnr,young2024nonequilibrium}, and glassiness~\cite{Vitelliglass}. 

Intuitively, non-reciprocity disrupts order and speeds up the dynamics, favoring fluid-like, disordered phases. 
This raises an interesting question: What effect does non-reciprocity have on many-body systems that are geometrically frustrated~\cite{introfrus}, and which, as a consequence, do not order even at arbitrarily low temperatures? 
Although analogies that suggest mappings between purely non-reciprocal and purely frustrated systems have recently been discussed~\cite{hanainonrecfrust,jorge2024active,Jorgeflows}, the interplay of non-reciprocal and frustrated interactions within the same system has hitherto remained unexplored. 

We consider a minimal model of two copies (species) of a square lattice Ising ($\mathbb{Z}_2$) gauge theory non-reciprocally coupled to each other, which we dub the non-reciprocal Ising gauge theory. 
Our choice is motivated by two factors: Firstly, recent theoretical work has shown a rich landscape of dynamical phenomena in out-of-equilibrium quantum~\cite{ChakrDFL2,SmithDFL,ChakrDFL1,KarpovDFL,ben2025many,halimeh2025quantum,osborne2024quantum,Banscar,gyawali2410observation} and classical~\cite{chakraborty2025fractional,homeier2025prethermal,raumonop,nisoli2021color} lattice gauge theories in two dimensions. 
Non-reciprocal counterparts of such gauge theories present a new frontier because broken detailed balance permits dynamics beyond global energy optimization. 
Secondly, recent theoretical and experimental activity has focused on realizing frustrated models in metamaterials and active matter~\cite{sirote2024emergent,coulais2016combinatorial,meeussen2020topological,merrigantop,jorge2024active,Jorgeflows,ortiz2016engineering}, 
which naturally host non-reciprocal interactions. 

Within Ising gauge theory, each individual species is characterised by topological order and associated fractionalized quasiparticle excitations. These quasiparticles are deconfined in two dimensions and, in the presence of Markovian stochastic dynamics, exhibit diffusive (i.e., random walk) behavior (see~\cite{supp}). 
In the presence of non-reciprocal interactions, we observe that the dynamics of the two species becomes correlated, generically leading to the confinement of inter-species quasiparticle pairs. This is seen in the asymptotic linear scaling of a corresponding Wilson loop observable, with a confinement length tuned by the non-reciprocal coupling strength. 
In the purely non-reciprocal (deconfined) regime, the interactions alter the diffusive behavior of the quasiparticles, leading at low temperatures to faster motion along self-avoiding trails on critical percolation clusters. 
When relating quasiparticle motion to magnetisation dynamics, we find that non-reciprocity speeds up the dynamics by relieving the topological logarithmic scaling of magnetic noise at intermediate times, followed by a slow down at the onset of a long-lived saturation plateau due to quasiparticle trapping. 
Our work demonstrates the general principles for how non-reciprocal interactions enrich the dynamical phenomenology of lattice gauge theories. 
%
%

\textbf{\textit{Model and observables --- }}
\begin{figure}
    \centering
    \includegraphics[scale = 0.21]{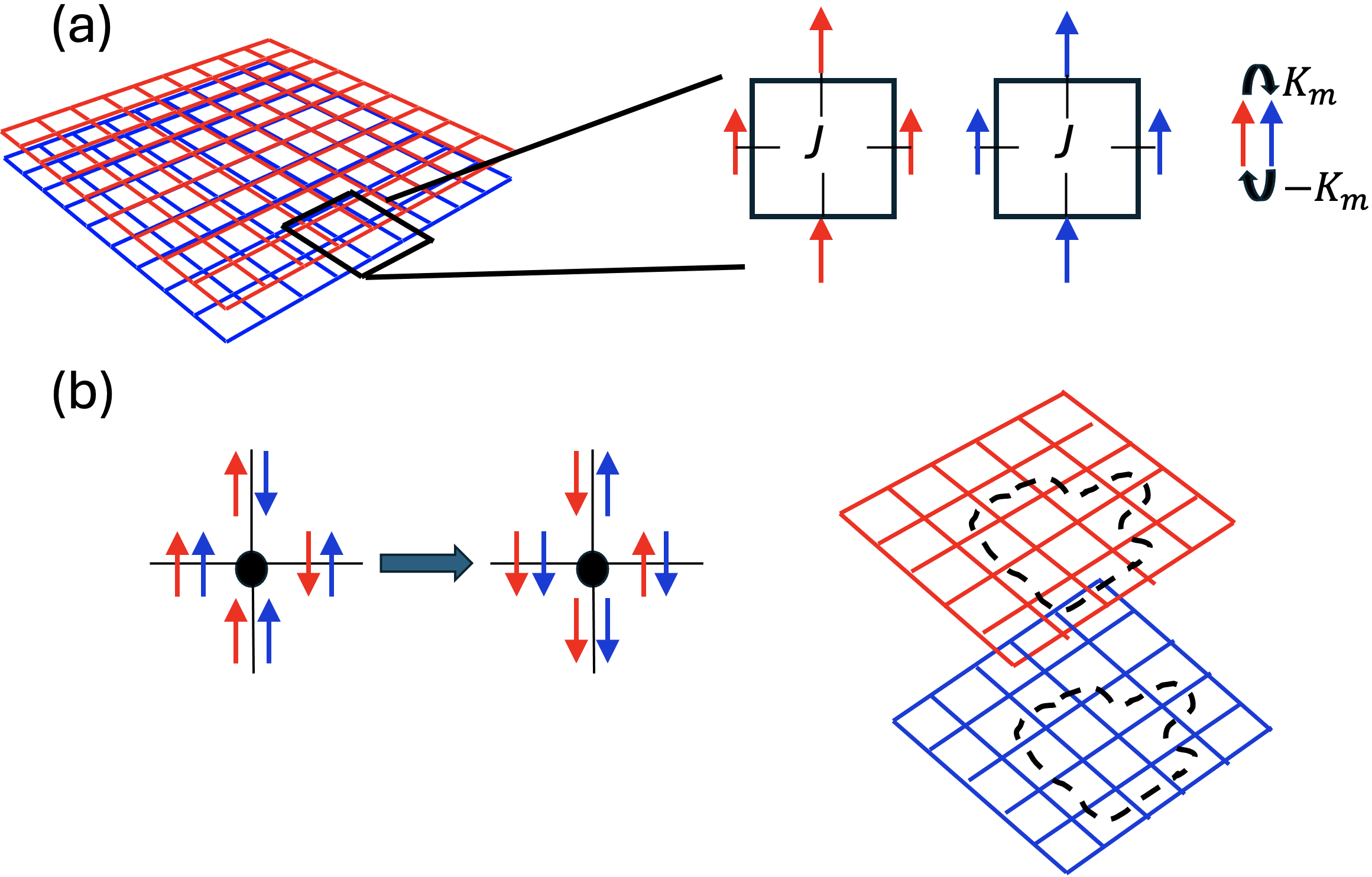}
    \caption{Non-reciprocal Ising gauge theory. (a) Two species ($A - \text{red}, B - \text{blue}$) of Ising variables ($\sigma^{A,B}_l$) live on the links $l$ of a square lattice, subject to a reciprocal intra-species plaquette interaction of strength $J$ and a nearest-neighbor inter-species interaction that has both reciprocal ($K_p$) and non-reciprocal ($K_m$) components. (b) The selfish energies are invariant under flipping all spins on the links connected to a pair of nearest-neighbor $A$ and $B$ sites. This extends to a general invariance of the simultaneous flip of all spins ($A,B$) along a closed loop on the dual lattice (black dashed lines).} 
    \label{fig1}
\end{figure}
Our model comprises two species, $A$ and $B$, of Ising variables on the links $l$ of a square lattice, $\sigma^{A,B}_l$. 
We consider intra-species four-body reciprocal interactions ($J$) around each square plaquette, characteristic of a $\mathbb{Z}_2$ gauge theory, and inter-species two-body onsite interactions that comprise reciprocal ($K_p$) and non-reciprocal ($K_m$) components (see~\cite{supp}). 
Non-reciprocal interactions invalidate the notion of a global energy, and we consider Monte-Carlo dynamics dictated by each species's selfish energy, $E^{A,B}$: 
\begin{equation}
\begin{aligned}
    E^A &= -J \sum_{\square_i, l_i \in \square_i} \sigma^{A}_{l_1} \sigma^{A}_{l_2} \sigma^{A}_{l_3} \sigma^{A}_{l_4} - (K_p+K_m) \sum_{l}
    \sigma^A_l \sigma^B_l \\
    E^B &= -J \sum_{\square_i, l_i \in \square_i} \sigma^{B}_{l_1} \sigma^{B}_{l_2} \sigma^{B}_{l_3} \sigma^{B}_{l_4} - (K_p-K_m) \sum_{l}
    \sigma^A_l \sigma^B_l 
\, , 
    \label{selfeneq}
\end{aligned}    
\end{equation}
where we assume without loss of generality that the coupling constants are non-negative (with $K_{p} = 0$ corresponding to the case of perfect non-reciprocity). 
The non-reciprocal interaction implies that each Ising variable $A$ prefers to align with its neighboring $B$, whereas $B$ prefers to anti-align with $A$. 
In the absence of non-reciprocal terms ($K_{p,m}=0$), this model has extensively many ground states corresponding to all configurations where the product of the four Ising variables around each plaquette is positive~\cite{wegner1971duality,Kogut-suss,Kogutrev}. 

The low-energy excitations take the typical form of point-like quasiparticles that are deconfined (namely, plaquettes where the product of the four spins is negative). Their motion is free across the lattice (when $K_{p,m}=0$) and quasiparticle dynamics mediates the low-energy re-arrangements of the spins, and thence the magnetization dynamics of the system (see~\cite{supp}). Non-vanishing inter-species couplings correlate the quasiparticle motion across the two lattices, as we uncover below. 

In the absence of conventional order and correlations, physical observables take the form of products of spins over closed contours $l_c$ along the links. These are the Wegner-Wilson (WW) loops $W^{A,B}$ for each species: 
\begin{equation}
W_A = \prod_{l \in L_c} \sigma^A_l 
\qquad 
W_B = \prod_{l \in L_c} \sigma^B_l 
\, , 
\label{wdef}
\end{equation}
with the combined loop $W_{AB} \equiv W_A W_B$. 
These observables are 
invariant under a local $\mathbb{Z}_2$ transformation corresponding to flipping all variables on links connected to a site, as illustrated in Fig.~\ref{fig1}(b).
In equilibrium, the expectation values of these observables decay with the length of the closed contour and distinguish between confined (linear scaling) and deconfined (quadratic scaling) phases of the gauge theory~\cite{wegner1971duality,Kogutrev}; however, in two-dimensional $\mathbb{Z}_2$ gauge theory only the deconfined phase occurs. 

We perform a numerical Monte Carlo study using single-spin-flip Glauber dynamics, with probability dependent on the selfish energy, 
\begin{equation}
    p(\sigma^i_l \rightarrow - \sigma^i_l) = \frac{1}{2}\bigg[1 - \text{tanh}\big(\dfrac{\Delta E^i}{2k_BT}\big) \bigg]
    \, , 
\end{equation}
where $\Delta E^i$ is the selfish energy change of species $i=A,B$, $k_B$ is the Boltzmann constant (which we set to $1$), and $T$ is the temperature. 
We start our simulations in a random initial state and let the system reach steady state before performing our measurements (see~EM).
%
%

\noindent \textbf{\textit{Interspecies quasiparticle pairing and confinement --- }}
\begin{figure}
     \centering
     \includegraphics[scale = 0.17]{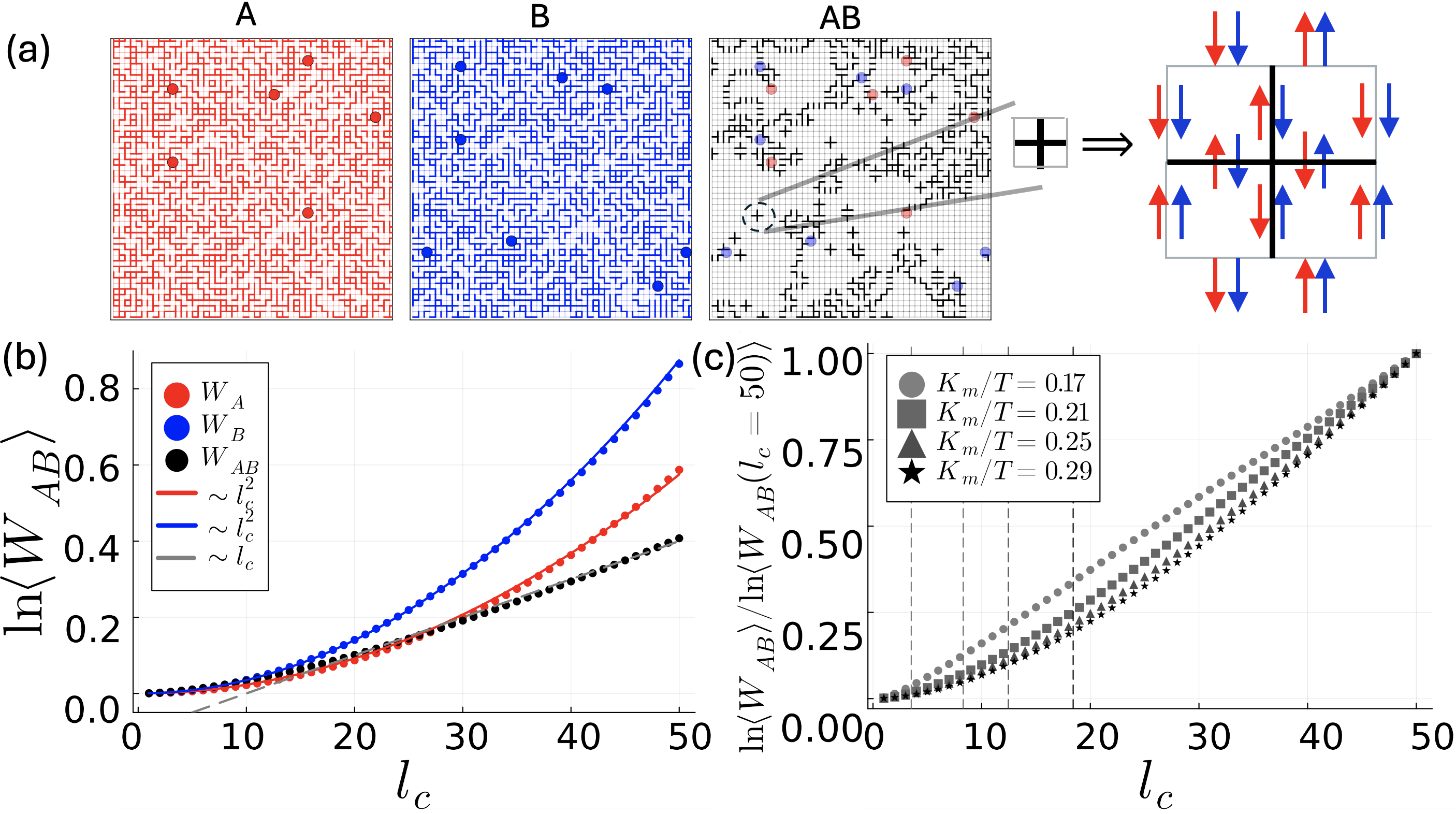}
     \caption{(a) Snapshots of $A$ (red) and $B$ (blue) spins, and their products $\sigma^A_i \sigma^B_i$ (black), with $-1$ values in bold, for $J/T = 3$, $K_p/T = 0.6$, $K_m/T = 0.35$ and $L = 50$.
     Quasiparticles are highlighted by filled circles. 
     (b) Time and history averages ($\langle \ldots \rangle$) of the WW observables. We see that $\langle W_{A,B} \rangle$ show the expected quadratic scaling for all distances (the quantitative difference is due to the different selfish inter-species couplings: $K_p \pm K_m$), whereas the average of their product $\langle W_{AB} \rangle$ displays a crossover from quadratic to linear scaling. (c) Dependence of $\langle W_{AB}(l_c) \rangle$ on $K_m$, with vertical dashed lines identifying the corresponding confinement length $l_{\rm cross}$. All WW curves are normalised by their value at $l_c = 50$, for ease of comparison. 
     The data in (b) and (c) are for $J/T = 4$, $K_p/T = 0.6$ and $L = 160$ and in (b) $K_m = 0.25$. 
     }
     \label{fig2}
\end{figure}
Let us consider the general case $K_p, \, K_m \neq 0$, and the behavior of the topological WW observables. At first sight the individual species $A,B$ appear unaffected (see Fig.~\ref{fig2}(b)), and indeed the time- and history-averaged quantity $-\ln\langle W_{A,B} \rangle$  exhibits the expected quadratic scaling. 
However, the configuration of the onsite products of $A$ and $B$ spin variables shows strong correlations, and the corresponding combined WW loop observables $\langle W_{AB} \rangle$ exhibit linear asymptotic scaling -- characteristic of confined excitations. 
Tuning the values of $K_{p,m}$ changes the crossover loop length $l_{\rm cross}$ from quadratic ($l_c < l_{\rm cross}$) to linear ($l_c > l_{\rm cross}$) scaling of $-\ln\langle W_{AB} \rangle$. 

This behavior can be understood by noting that the reciprocal coupling $K_p$ favours the ferromagnetic locking of species $A$ and $B$, causing the appearance of large regions where $\sigma^A \sigma^B = 1$, see Fig.~\ref{fig2}(a). Sparse antiferromagnetic $AB$ clusters are present, whose shape is controlled by the gauge correlations in the system. 
As a result, the quasiparticle motion becomes correlated. When a quasiparticle of one species traverses a ferromagnetic $AB$ region, it introduces a string of antiferromagnetic correlations between the two systems, which costs energy linear in the length of the path. This growing energy cost can be avoided by pairing $A$ and $B$ quasiparticles and moving them together across the ferromagnetic $AB$ region. We find that the system readily accomplishes this, thereby avoiding quasiparticle confinement \emph{within each species} and preserving the quadratic scaling of $W_{A,B}$. 

However, $A$-$B$ pairs of quasiparticles do not contribute to $W_{AB}$ in Eq.~\eqref{wdef}. The decay of the combined WW loop observable is controlled only by lone $A$ and $B$ quasiparticles, and therefore by the separation of $A$-$B$ pairs -- a process that is linearly confined by $K_p$ and by the corresponding presence of large ferromagnetic $AB$ regions. Asymptotically, $A$ and $B$ quasiparticles are bound to each other in pairs. The typical distance between $A$ and $B$ quasiparticles is controlled by the length scale set by the antiferromagnetic $AB$ clusters in Fig.~\ref{fig2}(a), namely their weighted diameter of gyration (see~EM). 
When the (combined) WW loops are smaller than this length, quasiparticle motion gives rise to quadratic scaling; when loops are larger than this scale, we observe linear scaling. 
If one increases $K_m$ at fixed $K_p$, the ferromagnetic regions reduce and the separation increases. 

While confinement per se is a result of the reciprocal coupling $K_p$, non-reciprocity allows for a unique tuneability at fixed temperature. As $K_m$ approaches zero, we find that $- \ln \langle W_{AB} \rangle$ exhibits purely linear scaling in the temperature range of interest, see Fig.~\ref{fig2}(c). 
%
%

\noindent \textbf{\textit{Quasiparticle motion and self-avoiding trails --- }}
\begin{figure}
    \centering
    \includegraphics[scale = 0.19]{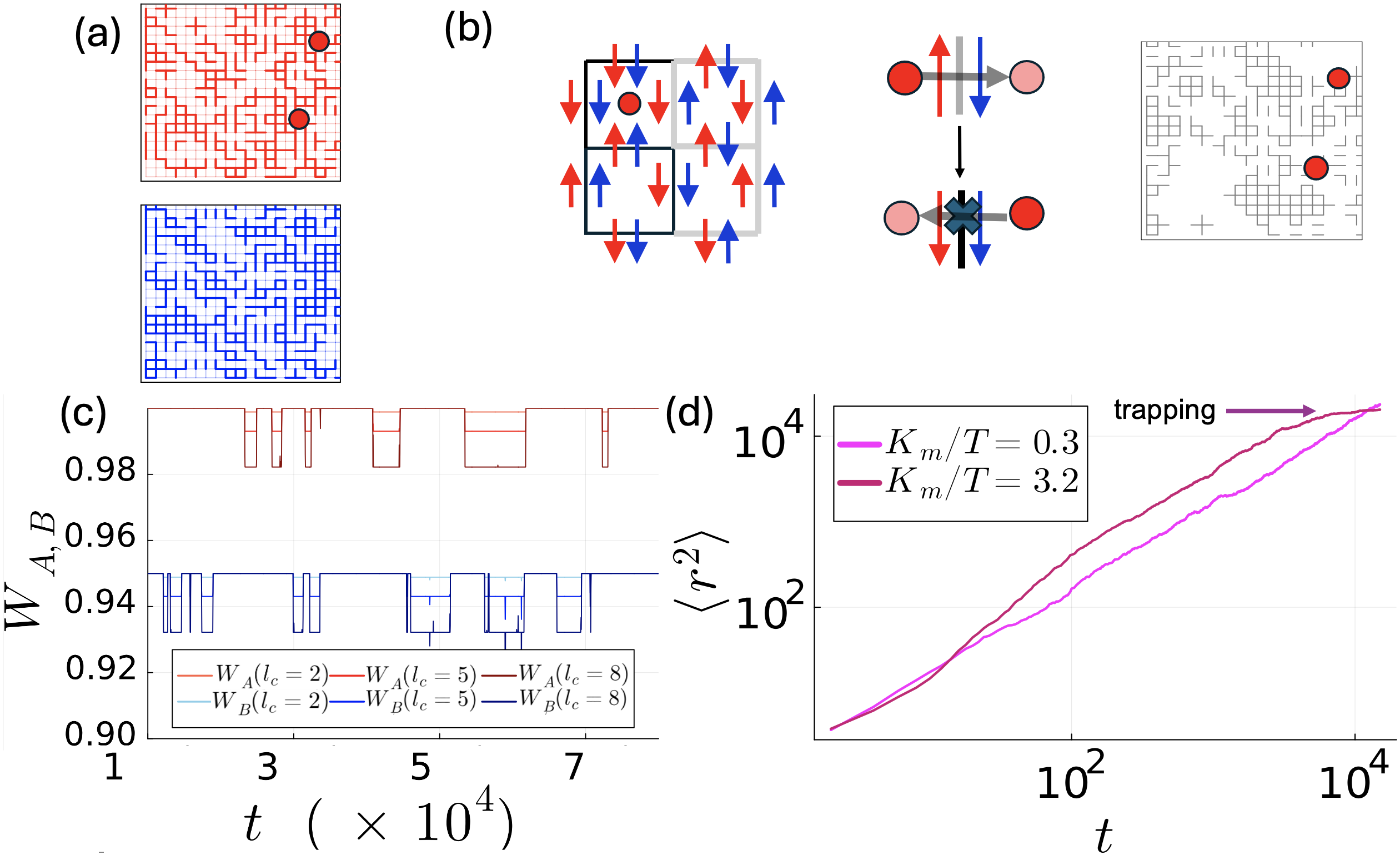}
    \caption{Quasiparticle dynamics in the low density regime ($J-K_m \gg T$). (a) A snapshot of the two species in one of the fluctuating windows, with two $A$ (red) quasiparticles and none in $B$ (blue). (b) Local snapshot showing the spins of both species near a single $A$ quasiparticle which hops upon flipping one of the $A$ spins around the excited plaquette. Strong non-reciprocal coupling biases such dynamics towards selfish-energy-lowering spin flips, resulting in the quasiparticles performing a self-avoiding trail (SAT) on a diluted lattice. (c) When $J-K_m \gg T$, finite size systems exhibit recurring time windows without excitations ($W_{A,B}=1, \, \forall l_c$ ) separated by fluctuating windows when quasiparticle pairs are created, propagate (deconfined, as evident from the quadratic length dependent scaling), and eventually annihilate within a species ($W_B$ is shifted by $0.05$ for visual clarity). Data shown for $L = 100$, $J/T = 6, K_m/T = 3.2$. (d) Mean squared displacement of a quasiparticle in the two-quasiparticle model for different values of $K_m/T$ with $L = 200$, averaged over $100$ trajectories. 
    The data contrast conventional RW behavior at small coupling with SAT behavior on a diluted lattice at strong coupling (see text and EM Fig.~\ref{fig5}). 
    }
    \label{fig3}
\end{figure}
Let us focus next on the case of purely non-reciprocal interactions ($K_p = 0$) and $J-K_m \gg T$. 
In this regime, we expect quasiparticle excitations to be sparse and deconfined, and events where they come close to one another across the two species are correspondingly rare. In our finite-size simulations, we typically find that quasiparticles are absent in one species when they are present in the other, and vice versa, as illustrated in Fig.~\ref{fig3}(c). 
We observe plateaux where the WW loop observables $W_A$ and $W_B$ saturate to $1$, corresponding to the absence of quasiparticles altogether (see~\cite{supp}). These are separated by intervals in time when either $W_A$ or $W_B$ fluctuate in time, corresponding to pair-creation, propagation, and annihilation of quasiparticles within one of the two species only. 
In this regime, the relaxation and magnetisation dynamics is controlled by single quasiparticle pair motion. 

The dynamics of these quasiparticles is controlled by non-reciprocity. Spins preferentially flip when this action lowers the inter-species interaction energy, and therefore the spin configuration of one species controls the motion of the quasiparticles in the other species. 
This effect becomes most pronounced in the strongly interacting regime $K_m \gg T$, when the bias for selfish-energy-decreasing spin-flip events is strong. The purely non-reciprocal nature of the interactions is fundamental to realise this regime: reciprocal inter-species interactions lead to the locking of one species onto the other, and possible effects similar to the ones discussed below show up only as transients rather than in steady state. By contrast, in the presence of purely non-reciprocal inter-species interactions, the motion of quasiparticles in one species favors one type of inter-species correlations (e.g., ferromagnetic) while the motion of quasiparticles in the other species favors the opposite inter-species correlations (e.g., antiferromagnetic). The two processes chase one another in perpetuity, enabling complex non-equilibrium steady-state dynamics. 

To explore this regime in detail, we focus on the motion of individual quasiparticles, and we restrict our simulations to the sector with two quasiparticles in species $A$ and none in species $B$, correspondingly constraining our Glauber dynamics not to create nor annihilate quasiparticles, see Fig.~\ref{fig3}(a,b). This is equivalent to studying a time window in Fig.~\ref{fig3}(c) where $W_A$ is fluctuating while $W_B$ remains saturated. If $K_m \gg T$, motion occurs preferentially via flips of spins $\sigma_l^A$ where $\sigma_l^A \sigma_l^B = -1$; flipping spins where $\sigma_l^A \sigma_l^B = +1$ is suppressed by the selfish energy Boltzmann factor $\sim e^{-2K_m/T} \ll 1$. As a result, the $A$ quasiparticles hop on a diluted lattice, illustrated in Fig.~\ref{fig3}(b), where the dilution is controlled by the spins of the other species ($B$). 
Given that spins in a $\mathbb{Z}_2$ lattice gauge are uncorrelated, and in the absence of reciprocal energy terms between the two species, it is reasonable to assume that the probability of a site having $\sigma_l^A \sigma_l^B = -1$ is $1/2$ and uniformly distributed, which means that the quasiparticles hop preferentially on a critical bond percolation cluster. 
By the same reasoning, quasiparticle motion is preferentially self-avoiding, since once a $\sigma_l^A$ spin is flipped, the $AB$ correlation changes from antiferromagnetic to ferromagnetic, and changing it back incurs a large selfish-energy cost. 
In summary, in the $K_m/T \to \infty$ limit, the quasiparticle performs a self-avoiding trail (SAT) on a critical percolation cluster. 

We show signatures of this behavior in Fig.~\ref{fig3}(d). For weak coupling, $K_m/T \ll 1$, we observe random walk (RW) behavior, $\langle r^2 \rangle \propto t$, as naively expected for $\mathbb{Z}_2$ gauge theory quasiparticles. At strong coupling, $K_m/T \gg 1$, we observe an initial upturn (transient superdiffusion), characteristic of SAT motion, followed by saturation at long times due to trapping in dead-ends, characteristic of SAT motion on a diluted network (see~EM). 
Of course, at very long times $t \gtrsim \exp(K_m/T)$, self-avoidance is violated, quasiparticles escape their traps, and ordinary diffusion resumes. 
The trapped quasiparticle regime is a long-lived metastable state induced by non-reciprocity (see EM and~\footnote{This long time trapping behavior is not observed in our simulations of the non-reciprocal Ising gauge theory because it pertains to quasiparticle pairs that separate to very large distances, which are only accessible for simulation times much longer than the ones considered in this study.}). 

%
%

\noindent \textbf{\textit{Topological logarithmic contribution to magnetisation dynamics --- }}
\begin{figure}
    \centering
    \includegraphics[scale = 0.17
    ]{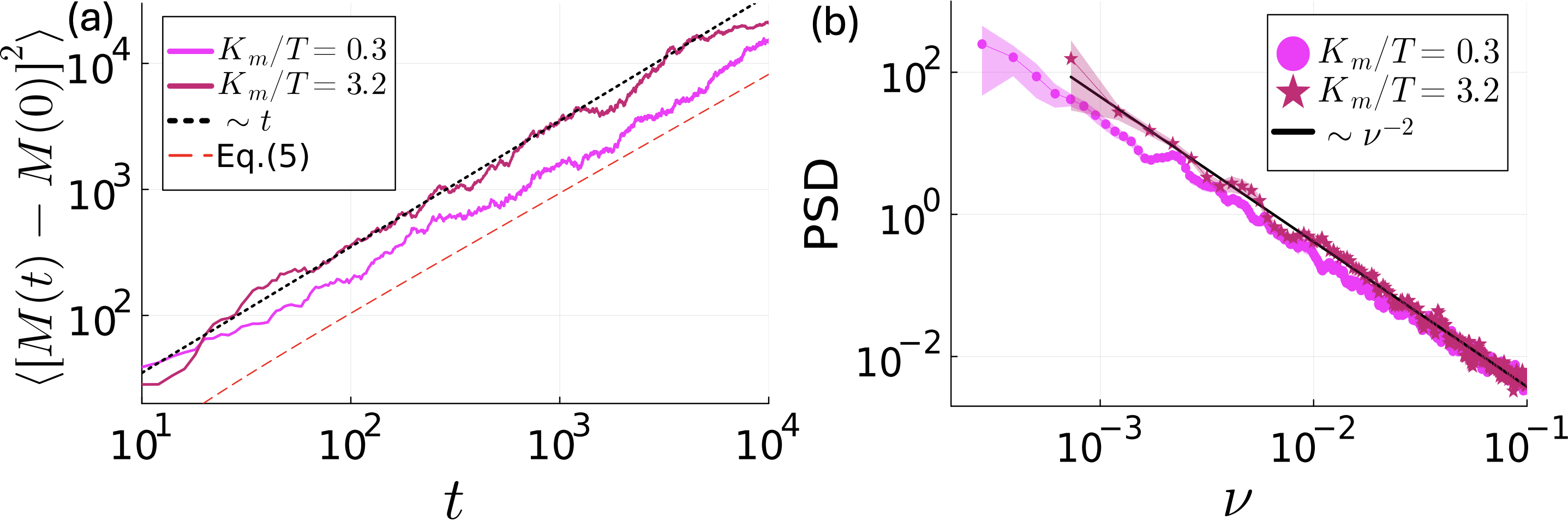}
    \caption{(a) Second moment of $M(t)$ for the two-quasiparticle model and different values of $K_m/T$ (weak and strong coupling). 
    The solid line is a guide to the eye showing linear $t$ scaling. 
    The dashed line corresponds to Eq.~\eqref{logmag} (no fitting parameters). 
    Inset: $\langle [ M(t) - M(0) ]^2 \rangle / t$ plotted in linear-log scale, to highlight the topological logarithmic contribution at weak coupling. 
    (b) Power spectral density of the magnetization averaged over the time windows when the WW of one species is fluctuating and that of the other is saturated in Fig.~\ref{fig3}(c). 
    We considered $J/T = 5$, $K_m/T = 0.3$ and $L = 120$; and $J/T = 6$, $K_m/T = 3.2$ and $L = 100$. We multiplied the $K_m/T = 0.3$ curve by $3.5$ for a better visual comparison. 
    The departure from inverse square scaling at weak coupling is due to topological logarithmic contributions that are lifted by the SAT behavior at strong coupling (see text).}
    \label{fig4}
\end{figure}
Quasiparticle motion controls the low-temperature spin dynamics, which in turn dictates the time evolution of the magnetization, $M(t)$. 
We proceed to investigate how non-reciprocity controls $M(t)$, which is the natural experimental observable. 

The relationship between quasiparticle motion and magnetisation dynamics in $\mathbb{Z}_2$ gauge theory is subtly complex (unlike the closely-resembling counterpart of $U(1)$ gauge theory in spin ice~\cite{udagawa2021spin}). 
Consider $M(t) = \sum_{l = 1}^{N} \sigma_l (t)$, where $l$ is the link index, $N$ is the total number of links, and $t$ labels Monte Carlo sweeps. 
With a single quasiparticle, $M(t)-M(0)$ is proportional to the sum of the spin values on the bonds traversed by the quasiparticle between time $t=0$ and time $t$. 
Since a spin flips upon being traversed, all spins traversed an even number of times cancel out, and therefore 
\begin{equation}
    M(t)-M(0) = -2 \sum_{l \in {\rm odd}_t} \sigma_l (0)
    \, , 
    \label{eqMt}
\end{equation}
where ${\rm odd}_t$ is the set of all the bonds traversed an odd number of times between time $t=0$ and time $t$. 

Given that Ising gauge theory has no correlations other than topological ones (see~\cite{supp}), the magnetization dynamics of our system in the vanishing quasiparticle density limit can be largely recast as that of an Ising paramagnet whose spins on the bonds of the lattice are flipped in time by a random walker. This deceivingly simple problem has very interesting dynamical signatures. 

Using established results on repeated bond traversals of two-dimensional random walks~\cite{antal2002repeated}, one can analytically calculate the asymptotic scaling ($t \gg 1$) of the expectation value of the second moment of $M(t) - M(0)$ to get (see~EM) 
\begin{equation}
   \langle [M(t \gg 1) - M(0)]^2 \rangle  =  \dfrac{4 \pi t}{\text{log}(8t)} \bigg[1 - \dfrac{\frac{3\pi}{4} - \frac{3}{2} (1-\gamma)}{\text{log}(8t)}  \bigg] + \, ... 
   \, , 
\label{logmag}
\end{equation}
where the dotted terms are of $O(t \, \text{log}^{-3} (8t))$ and $\gamma \simeq 0.57721$ is Euler's constant. 
Correspondingly, the power spectral density (PSD) exhibits a low-frequency scaling with leading term $\mathrm{PSD}(\nu) \sim A_1/(\nu^2 \text{log}(A_2/\nu))$, with constants $A_1$ and $A_2$. 
In our system, the logarithmic contribution is topological, arising from the fractionalized point-like excitations of $\mathbb{Z}_2$ gauge theory and recurrence properties of two-dimensional random walks. It persists for weak non-reciprocal coupling as shown in Fig.~\ref{fig4}(a,b).

As the coupling increases, quasiparticle hops that retrace a bond become less likely. Asymptotically for $K_m/T \gg 1$, the quasiparticles perform SATs and $M(t)-M(0)$ reduces to a sum of random $\pm 1$ increments --- equivalent to a 1D RW process with scaling $\langle [M(t)-M(0)]^2\rangle \sim t$ and $\mathrm{PSD}(\nu) \sim \nu^{-2}$, once again in agreement with our numerics in Fig.~\ref{fig4}(a,b). 
Strong coupling also dilutes the network over which the SAT takes place, and eventually all trajectories terminate in (metastable) dead ends at long times. This behavior results in an eventual departure from diffusive scaling, and in the second moment saturating beyond the trapping time, as illustrated in Fig.~\ref{fig4}(a) (see~EM). 
%
%

\noindent \textbf{\textit{Discussion --- }}
Non-reciprocity and geometric frustration are two distinct pathways to induce complex far-from-equilibrium behavior and unconventional order in many-body systems. Our results illustrate the rich phenomenology that derives from their interplay, paving the way to the field of frustrated non-reciprocity and non-reciprocal many-body topological phases. 

We envision generalizations to a broad classes of systems, for example a three-dimensional non-reciprocal Ising gauge theory, where the reciprocal counterpart exhibits a finite temperature phase transition between a confined and a deconfined phase in equilibrium. Another example are non-reciprocal $U(1)$ models, combining non-reciprocal interactions and spin ice physics, exploring the effects on monopole dynamics~\cite{chakraborty2025fractional,hallen2022dynamical,castelnovoreview} and glassiness~\cite{TCMglass}.
Quantum $\mathbb{Z}_2$ gauge theories have led to numerous breakthroughs in strongly correlated many body systems~\cite{moessner2021topological}, and their non-reciprocal counterparts will provide further richness. Recent work has highlighted the possibility of realizing non-reciprocal interactions between two copies of all-to-all coupled quantum spins in open systems~\cite{brunelliquantnon}, and we hope for future extensions to short-ranged frustrated quantum systems. 
%
%

\noindent \textbf{\textit{Acknowledgements --- }}
We thank Lilian Dahan for collaboration on related work. We also thank Marin Bukov, Michael Knap, Johannes Knolle, Paul Krapivsky and Peter Littlewood for valuable discussions. 
%
%

%
%
\clearpage
\newpage
\clearpage
\section*{End Matter}


\vspace{1em}
\section{Numerical details}
We perform Glauber dynamics simulations for Ising spins (of species $A$ and $B$) living on the bonds of an $L \times L$ square lattice, with time unit defined as $4L^2$ attempted updates of randomly selected spins. All our simulations use periodic boundary conditions, starting from random initial conditions for both species. 

For the WW observables in Eq.~\eqref{wdef}, we choose every site of the lattice as the bottom left corner of an $l_c \times l_c$ square, where the edges of the square run along the links of the lattice. We then average over all such possible squares of linear size $l_c$. We measure the WW observables every $50$ units of time and we take the spatiotemporal average (after the system has reached steady state. In order to establish that the system has reached steady state, we initially measure the WW observables as a function of time and monitor their drift. Once this is no longer statistically significant, we assume that steady state has been reached. We have checked that increasing the measurement time does not change the data in Fig.~ \ref{fig2}(b,c)). 

To obtain the crossover length in the inset of Fig.~ \ref{fig2}(c) we use the following procedure. We perform a parabolic fit of progressively higher data points, starting from the leftmost three points, until the goodness of fit falls below a chosen threshold ($R^2 = 0.85$). We then do the same using a linear fit, starting from the rightmost three values and then adding lower data points until reaching the goodness of fit threshold. The crossover length is defined via slope matching, i.e, the point of the last parabola in the fitting procedure above whose slope matches that of the last linear fit. 

\begin{figure}
    \centering
    \includegraphics[width=0.7\linewidth]{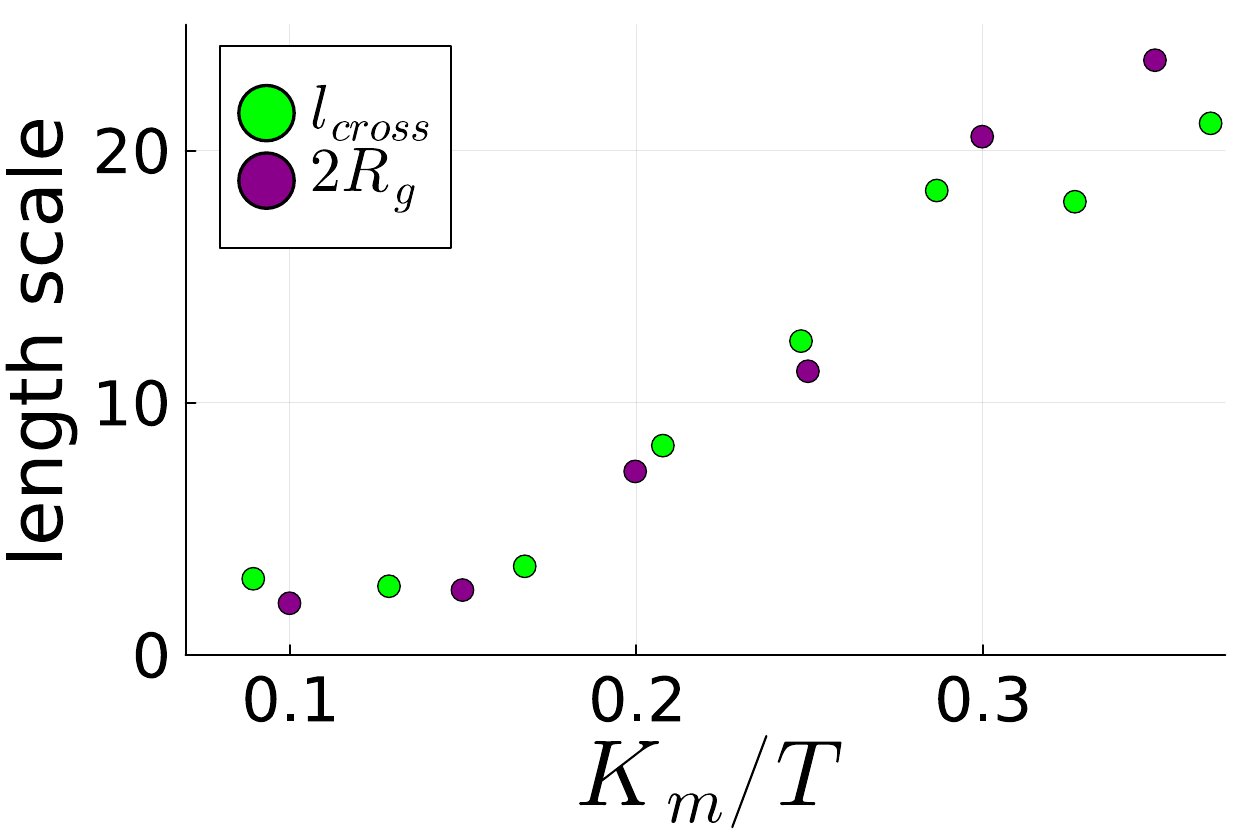}
    \caption{The confinement length scale (linear to quadratic crossover) from the $W_{AB}$ curves in Fig.~\ref{fig2}(c), compared to the diameter of gyration of the antiferromagnetic clusters of $\sigma^A_l \sigma^B_l = -1$, for different values of $K_m/T$.}
    \label{Figs/figlcross}
\end{figure}

To obtain the weighted diameter of gyration (physically relevant length scale) of the clusters of connected $\sigma^A \sigma^B = -1$ variables (see Fig.~\ref{fig2}(b) for reference), we compute the radius of gyration for each cluster defined as the pairwise average distance between its sites. Finally, we average these values weighted by the cluster size, and multiply by two to obtain the diameter. 
As shown in Fig.~\ref{Figs/figlcross}, we find good agreement between the diameter of gyration and the confinement length scale as function of $K_m/T$.

For the data in Fig.~\ref{fig3}(d) and~\ref{fig4}(a), we start with a random configuration in the $A$ and $B$ copies, with 2 quasiparticles in $A$ and none in $B$. We then constrain our Glauber dynamics not to create nor annihilate quasiparticles. Our curves are averages over $100$ histories for each value of the simulation parameters. 

For the PSD data in Fig.~\ref{fig4}(b), we identify each window where $W_{A,B}$ is fluctuating while $W_{B,A}$ remains saturated, and extract the time-dependent magnetisation from each window. We discard the first $100$ time data points, as well as all time points beyond the middle of the time window, to discount the diffusive behavior right after creation and close to annihilation. We then compute the PSD and take the average over many windows using the frequency bin size of the largest dynamical window. (This naturally results in progressively larger statistical error bars for lower frequency data points, as shown by the shaded region in Fig.~\ref{fig4}(b)). 
%
%

\section{Toy model of magnetization dynamics}
\begin{figure}
    \centering
    \includegraphics[scale = 0.18]{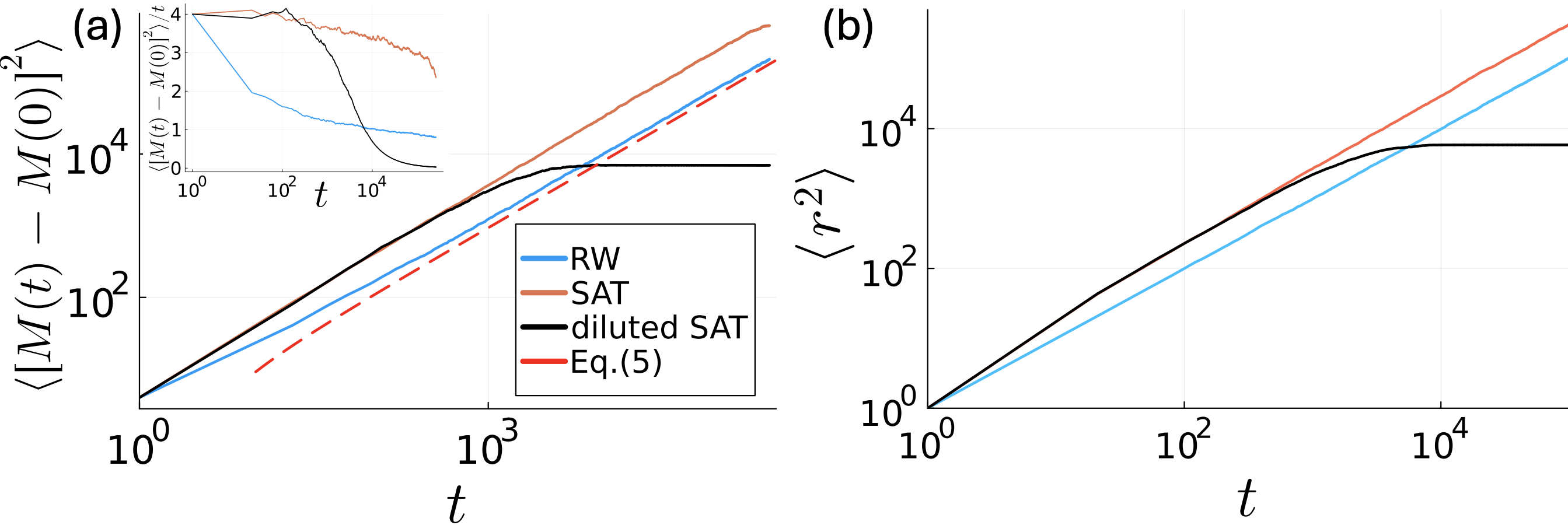}
    \caption{Magnetization and particle dynamics for the toy model. (a) Second moment of the magnetization for the ordinary RW, for the SAT, and for the SAT on a critical percolation lattice. The red solid line shows the analytical result in Eq.~\eqref{logmag}. Inset: Same data divided by $t$ to highlight the log contribution in the ordinary RW model, and the approximately linear scaling in the SAT (up to the trapping time). (b) Mean squared displacement for the same models.}
    \label{fig5}
\end{figure}
In the main text, we presented numerical and analytical evidence for the qualitatively different motion of the quasiparticles in the weak and strong non-reciprocal coupling regimes. In this section we compare those results to a toy model of paramagnetic Ising spins on the links of a square lattice, subject to stochastic dynamics in the form of a random walker on the sites of the lattice, whereby a spin is flipped upon being traversed by the walker. We contrast ordinary RW motion with SAT motion, and with SAT motion on a critical percolation cluster. We use the same unit of time as in the main text. 

The mean-squared displacement of the walker is shown in Fig.~\ref{fig5}(b). Ordinary RW motion shows linear scaling, as expected. On the contrary, SAT on critical percolation clusters displays qualitatively similar behavior to that in Fig.~\ref{fig3}(d) in the main text: transient superdiffusivity at short times and eventual saturation at long times due to trapping in dead-ends~\footnote{In this work, we were unable to find conclusive numerical evidence that the diluted network is indeed a critical percolation cluster. We note that the trapping time and the $\langle r^2 \rangle$ saturation are properties of the network and do not depend on the value of $K_m/T$; therefore it may be possible to infer the nature of the network from these values -- a task that is left for future work.}. (For comparison, we also show the SAT case without dilution, which does not saturate.) 

\begin{figure}
    \centering
    \includegraphics[scale = 0.22]{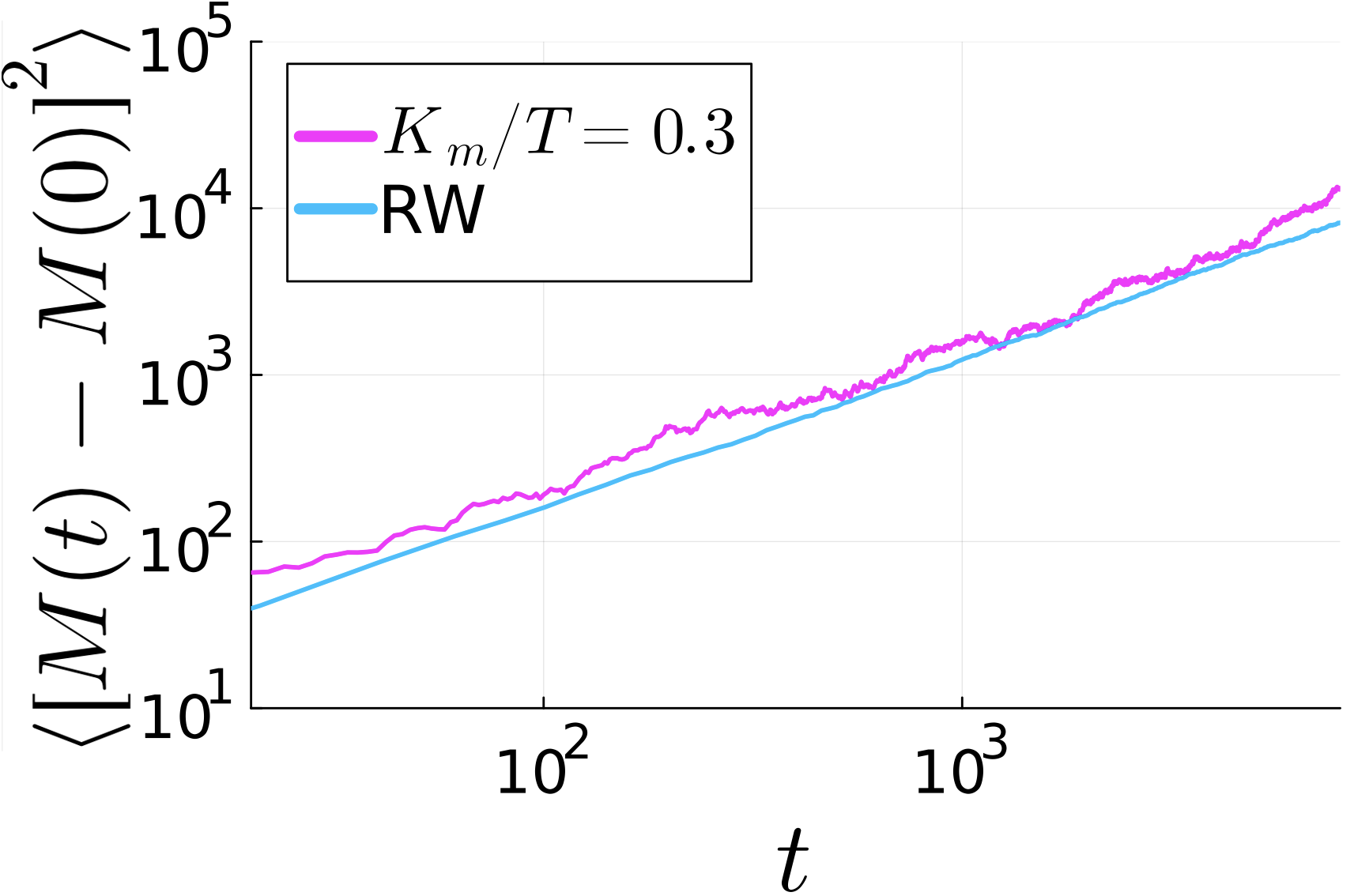}
    \caption{Weak coupling magnetization dynamics of the non-reciprocal Ising gauge theory and that of the random walk toy model (comparing the data in Fig.~\ref{fig4}(a) and in Fig.~\ref{fig5}(a)).}
    \label{fig6}
\end{figure}

The second moment of the magnetization is shown in Fig.~\ref{fig5}(a). Ordinary RW motion shows departure from linear scaling, asymptotically approaching our analytical result in Eq.~\eqref{logmag}, as also seen in Fig.~\ref{fig4}(a) for weak non-reciprocal coupling in the main text.
The analytical asymptotic behavior only onsets after long times, $t > 10^4$; our toy model however shows good agreement with the weak-non-reciprocal behavior in the main text already at short times (Fig.~\ref{fig6}).

On the other hand, SAT, corresponding to strong non-reciprocal coupling in the main text, lifts the logarithmic contribution and restores linear in $t$ behavior (up to the trapping time). 
%
%

\section{Quasiparticle and magnetisation dynamics}
Here we provide a detailed calculation of the expectation value of the second moment of $M(t)-M(0) = -2 \sum_{l \in \rm{odd}_t} \sigma_l (0)$, for the square lattice Ising paramagnet with RW stochastic dynamics introduced in the previous section. After squaring and taking the average over random initial conditions, we obtain 
\begin{equation}
  \langle [M(t)-M(0)]^2 \rangle  = 4 \langle \text{no. of oddly-flipped bonds} \rangle_t
\, . 
\end{equation}
This is because the cross terms $\sigma_l(0) \sigma_{l'}(0)$ with $l \neq l'$ average to $0$ (assuming as appropriate for an Ising gauge theory that there are no spatial correlations between the spins), and the diagonal terms square to $1$ for every bond flipped an odd number of times between $t=0$ and $t$, i.e., $\forall l \in \rm{odd}_t$. The problem reduces to finding the average number of bonds that an ordinary two-dimensional RW traverses an odd number of times in time $t$. 

To calculate this, we use the result obtained in Ref.~\onlinecite{antal2002repeated}, which found that the average number of bonds crossed $m$ times in a given interval $t$ is 
\begin{equation}
    B_m (t) = \dfrac{4 \pi^2 t}{\text{log}^2(8t)}e^{-\mu} \bigg[1 + (\mu-2) \dfrac{\pi/2 + 1 - \gamma}{log(8t)} + \text{O}(\text{log}^{-2}(8t)) \bigg]
\, , 
\label{B_meq}
\end{equation}
at long times, where $\mu = 2\pi m/\text{log}(8t)$ and $\gamma$ is Euler's constant. 

Let us first calculate the fraction of bonds traversed by the random walk that are crossed an odd number of times.
We define 
\begin{equation}
    F(t \gg 1) = \dfrac{\sum_{m = 1,3,5...} B_m (t \gg 1)}{\sum_{m' = 1,2,3,4,...} B_{m'} (t \gg 1)}
\, . 
\end{equation}
\begin{figure}[h!]
    \centering
    \includegraphics[scale = 0.25]{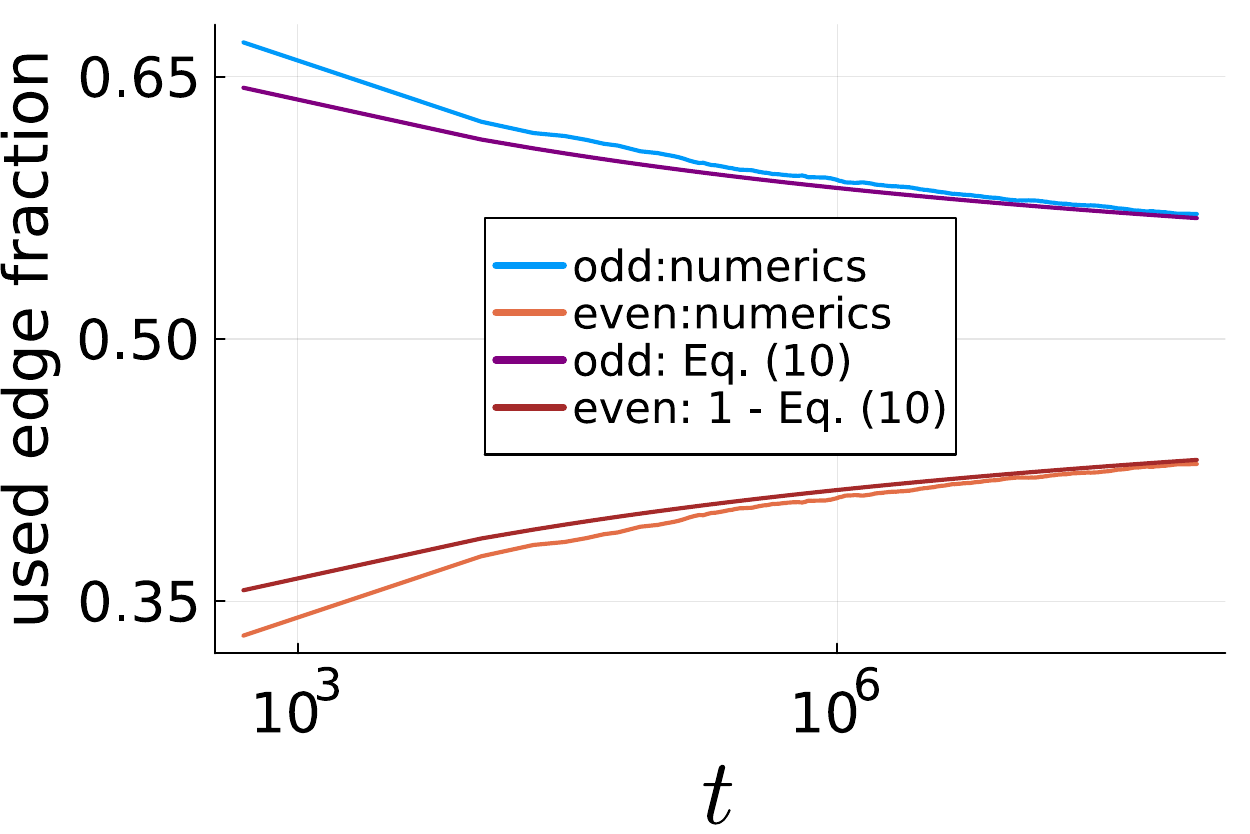}
    \caption{Fraction of bonds traversed an odd (even) number of times by a two-dimensional random walk. Numerical (orange+green) and analytical (blue+purple) results agree asymptotically.}
    \label{fig7}
\end{figure}
After some algebra one can write the above expression in the form
\begin{equation}
    F(t) = \dfrac{C_1(1-q^2)+C_2(1+q^2)}{(C_1(1-q)+C_2)(1+q)^2}
\, , 
\end{equation}
where $q = e^{-\mu}$, $C_1 = 1-2(\frac{\pi}{2}+1-\gamma)/\text{log}(8t)$ and $C_2 = 2 \pi (\frac{\pi}{2}+1-\gamma) / \text{log}(8t)$, and we used Eq.~\eqref{B_meq} only up to the O($t\text{log}^{-3}(8t)$) term. 
In the desired large $t$ limit, this gives
\begin{equation}
    F(t) = \dfrac{1}{2} + \dfrac{\pi}{2\text{log}(8t)} - \dfrac{\pi}{\text{log}^2(8t)} + \text{O}(\text{log}^{-3}(8t))
\, . 
\end{equation}
We have checked that this estimate agrees asymptotically with our numerical results, as shown in Fig.~\ref{fig7}. 
We see that the fraction of oddly traversed bonds approaches $1/2$ only very slowly due to the logarithmic contribution. 

In Ref.~\onlinecite{antal2002repeated}, the authors obtained the average number of distinct bonds traversed by the walker at time $t$ as $\beta(t) = 2\pi t/\text{log}(8t) - (7 \pi^2/2 + 3\pi (1-\gamma))t/(\text{log}^2 (8t))+ \text{O}(t\text{log}^{-3}(8t))$. Using this result one can obtain the average number of bonds crossed an odd number of times in time $t$, and from it obtain the result mentioned in the main text in Eq.~\eqref{eqMt}: $4 F(t) \beta(t)$. We see in Fig.~\ref{fig5}(b) that the asymptotic expression agrees asymptotically well with numerics, without any fitting parameters. 
%
%

\bibliographystyle{apsrev4-2}
\bibliography{Ref}

@book{introfrus,
  title = {Introduction to Frustrated Magnetism: Materials,  Experiments,  Theory},
  ISBN = {9783642105890},
  ISSN = {0171-1873},
  url = {http://dx.doi.org/10.1007/978-3-642-10589-0},
  DOI = {10.1007/978-3-642-10589-0},
  journal = {Springer Series in Solid-State Sciences},
  publisher = {Springer Berlin Heidelberg},
  year = {2011}
}

@article{fruchartreview,
  title={Nonreciprocal many-body physics},
  author={Fruchart, Michel and Vitelli, Vincenzo},
  journal={arXiv:2602.11111},
  year={2026},
  url = {https://arxiv.org/abs/2602.11111}
}

@article{young2024nonequilibrium,
  title={Nonequilibrium universality of the nonreciprocally coupled O (n1)$\times$ O (n2) model},
  author={Young, JT and Gorshkov, AV and Maghrebi, M},
  journal={arXiv:2411.12680},
  year={2024},
  url = {https://arxiv.org/abs/2411.12680}
}

@article{Ivlevnr,
  title = {Statistical Mechanics where Newton's Third Law is Broken},
  author = {Ivlev, A. V. and Bartnick, J. and Heinen, M. and Du, C.-R. and Nosenko, V. and L\"owen, H.},
  journal = {Phys. Rev. X},
  volume = {5},
  issue = {1},
  pages = {011035},
  numpages = {10},
  year = {2015},
  month = {Mar},
  publisher = {American Physical Society},
  doi = {10.1103/PhysRevX.5.011035},
  url = {https://link.aps.org/doi/10.1103/PhysRevX.5.011035}
}

@article{supriyonr,
  title = {Scalar Active Mixtures: The Nonreciprocal Cahn-Hilliard Model},
  author = {Saha, Suropriya and Agudo-Canalejo, Jaime and Golestanian, Ramin},
  journal = {Phys. Rev. X},
  volume = {10},
  issue = {4},
  pages = {041009},
  numpages = {14},
  year = {2020},
  month = {Oct},
  publisher = {American Physical Society},
  doi = {10.1103/PhysRevX.10.041009},
  url = {https://link.aps.org/doi/10.1103/PhysRevX.10.041009}
}

@article{Duannr,
  title = {Phase coexistence in nonreciprocal quorum-sensing active matter},
  author = {Duan, Yu and Agudo-Canalejo, Jaime and Golestanian, Ramin and Mahault, Beno\^{\i}t},
  journal = {Phys. Rev. Res.},
  volume = {7},
  issue = {1},
  pages = {013234},
  numpages = {21},
  year = {2025},
  month = {Mar},
  publisher = {American Physical Society},
  doi = {10.1103/PhysRevResearch.7.013234},
  url = {https://link.aps.org/doi/10.1103/PhysRevResearch.7.013234}
}

@article{you2020nonreciprocity,
  title={Nonreciprocity as a generic route to traveling states},
  author={You, Zhihong and Baskaran, Aparna and Marchetti, M Cristina},
  journal={Proceedings of the National Academy of Sciences},
  volume={117},
  number={33},
  pages={19767--19772},
  year={2020},
  publisher={National Academy of Sciences},
  url = {https://www.pnas.org/doi/abs/10.1073/pnas.2010318117}
}

@article{NRIsing,
  title = {Nonreciprocal Ising Model},
  author = {Avni, Yael and Fruchart, Michel and Martin, David and Seara, Daniel and Vitelli, Vincenzo},
  journal = {Phys. Rev. Lett.},
  volume = {134},
  issue = {11},
  pages = {117103},
  numpages = {9},
  year = {2025},
  month = {Mar},
  publisher = {American Physical Society},
  doi = {10.1103/PhysRevLett.134.117103},
  url = {https://link.aps.org/doi/10.1103/PhysRevLett.134.117103}
}

@article{DynphaseNRIsing,
  title = {Dynamical phase transitions in the nonreciprocal Ising model},
  author = {Avni, Yael and Fruchart, Michel and Martin, David and Seara, Daniel and Vitelli, Vincenzo},
  journal = {Phys. Rev. E},
  volume = {111},
  issue = {3},
  pages = {034124},
  numpages = {41},
  year = {2025},
  month = {Mar},
  publisher = {American Physical Society},
  doi = {10.1103/PhysRevE.111.034124},
  url = {https://link.aps.org/doi/10.1103/PhysRevE.111.034124}
}

@article{derrida1987exactly,
  title={An exactly solvable asymmetric neural network model},
  author={Derrida, Bernard and Gardner, Elizabeth and Zippelius, Anne},
  journal={Europhysics Letters},
  volume={4},
  number={2},
  pages={167},
  year={1987},
  publisher={IOP Publishing},
url = {https://iopscience.iop.org/article/10.1209/0295-5075/4/2/007/meta?casa_token=-w-995oq4N8AAAAA:-YCRJXSIIcbYgJRN9zb0I6tP4v39ztO2xdt9XF9SVX7XQ1iczKuIgeBMmh9paFfaXnkhtsuNlABRf7qxvmniPplcMbnv}
}

@article{parisi1986asymmetric,
  title={Asymmetric neural networks and the process of learning},
  author={Parisi, Giorgio},
  journal={Journal of Physics A: Mathematical and General},
  volume={19},
  number={11},
  pages={L675},
  year={1986},
  publisher={IOP Publishing},
url = {https://iopscience.iop.org/article/10.1088/0305-4470/19/11/005/meta?casa_token=yDyZq7UMklYAAAAA:-QJI52yRz1zfJ13CQkWmfEq69shNcIHdvh0LuSuAGug6I9cnq-I7Ntg3N8KYhCFuwl8Z15faqqijtjOu9qlbUkwpx-eW}
}

@article{ortiz2016engineering,
  title={Engineering of frustration in colloidal artificial ices realized on microfeatured grooved lattices},
  author={Ortiz-Ambriz, Antonio and Tierno, Pietro},
  journal={Nature communications},
  volume={7},
  number={1},
  pages={10575},
  year={2016},
  publisher={Nature Publishing Group UK London},
  url = {https://www.nature.com/articles/ncomms10575}
}

@article{brunelliquantnon,
  title = {Nonreciprocal Synchronization of Active Quantum Spins},
  author = {Nadolny, Tobias and Bruder, Christoph and Brunelli, Matteo},
  journal = {Phys. Rev. X},
  volume = {15},
  issue = {1},
  pages = {011010},
  numpages = {21},
  year = {2025},
  month = {Jan},
  publisher = {American Physical Society},
  doi = {10.1103/PhysRevX.15.011010},
  url = {https://link.aps.org/doi/10.1103/PhysRevX.15.011010}
}

@article{asymmneur,
  title = {Temporal Association in Asymmetric Neural Networks},
  author = {Sompolinsky, H. and Kanter, I.},
  journal = {Phys. Rev. Lett.},
  volume = {57},
  issue = {22},
  pages = {2861--2864},
  numpages = {0},
  year = {1986},
  month = {Dec},
  publisher = {American Physical Society},
  doi = {10.1103/PhysRevLett.57.2861},
  url = {https://link.aps.org/doi/10.1103/PhysRevLett.57.2861}
}

@article{hongstrogprl,
  title = {Kuramoto Model of Coupled Oscillators with Positive and Negative Coupling Parameters: An Example of Conformist and Contrarian Oscillators},
  author = {Hong, Hyunsuk and Strogatz, Steven H.},
  journal = {Phys. Rev. Lett.},
  volume = {106},
  issue = {5},
  pages = {054102},
  numpages = {4},
  year = {2011},
  month = {Feb},
  publisher = {American Physical Society},
  doi = {10.1103/PhysRevLett.106.054102},
  url = {https://link.aps.org/doi/10.1103/PhysRevLett.106.054102}
}

@article{strog2,
  title = {Conformists and contrarians in a Kuramoto model with identical natural frequencies},
  author = {Hong, Hyunsuk and Strogatz, Steven H.},
  journal = {Phys. Rev. E},
  volume = {84},
  issue = {4},
  pages = {046202},
  numpages = {6},
  year = {2011},
  month = {Oct},
  publisher = {American Physical Society},
  doi = {10.1103/PhysRevE.84.046202},
  url = {https://link.aps.org/doi/10.1103/PhysRevE.84.046202}
}

@article{clerk2022introduction,
  title={Introduction to quantum non-reciprocal interactions: from non-Hermitian Hamiltonians to quantum master equations and quantum feedforward schemes},
  author={Clerk, Aashish},
  journal={SciPost Physics Lecture Notes},
  pages={044},
  year={2022},
url = {https://www.scipost.org/SciPostPhysLectNotes.44?acad_field_slug=politicalscience}
}

@article{lorenzana2025nonreciprocity,
  title={When is nonreciprocity relevant?},
  author={Lorenzana, Giulia Garcia and Martin, David and Avni, Yael and Seara, Daniel S and Fruchart, Michel and Biroli, Giulio and Vitelli, Vincenzo},
  journal={arXiv:2509.17972},
  year={2025},
  url = {https://arxiv.org/abs/2509.17972}
}

@article{fruchart2021non,
  title={Non-reciprocal phase transitions},
  author={Fruchart, Michel and Hanai, Ryo and Littlewood, Peter B and Vitelli, Vincenzo},
  journal={Nature},
  volume={592},
  number={7854},
  pages={363--369},
  year={2021},
  publisher={Nature Publishing Group UK London},
url  = {https://www.nature.com/articles/s41586-021-03375-9}
}

@article{hanainonrecfrust,
  title = {Nonreciprocal Frustration: Time Crystalline Order-by-Disorder Phenomenon and a Spin-Glass-like State},
  author = {Hanai, Ryo},
  journal = {Phys. Rev. X},
  volume = {14},
  issue = {1},
  pages = {011029},
  numpages = {25},
  year = {2024},
  month = {Feb},
  publisher = {American Physical Society},
  doi = {10.1103/PhysRevX.14.011029},
  url = {https://link.aps.org/doi/10.1103/PhysRevX.14.011029}
}

@article{bascompte2006asymmetric,
  title={Asymmetric coevolutionary networks facilitate biodiversity maintenance},
  author={Bascompte, Jordi and Jordano, Pedro and Olesen, Jens M},
  journal={Science},
  volume={312},
  number={5772},
  pages={431--433},
  year={2006},
  publisher={American Association for the Advancement of Science},
url = {https://www.science.org/doi/full/10.1126/science.1123412?casa_token=VfxyegGxotwAAAAA%3AYD2vm_mj4keDn-91qevl-KoBAm98_gRNqKzS9MbwcHGip95Wndu-HJQay31MdL2fCp3d1Ga4NVUVIEo}
}

@article{gamenonrec,
  title = {Unlearnable Games and ``Satisficing'' Decisions: A Simple Model for a Complex World},
  author = {Garnier-Brun, J\'er\^ome and Benzaquen, Michael and Bouchaud, Jean-Philippe},
  journal = {Phys. Rev. X},
  volume = {14},
  issue = {2},
  pages = {021039},
  numpages = {38},
  year = {2024},
  month = {Jun},
  publisher = {American Physical Society},
  doi = {10.1103/PhysRevX.14.021039},
  url = {https://link.aps.org/doi/10.1103/PhysRevX.14.021039}
}

@article{strog1,
  title = {Kuramoto Model of Coupled Oscillators with Positive and Negative Coupling Parameters: An Example of Conformist and Contrarian Oscillators},
  author = {Hong, Hyunsuk and Strogatz, Steven H.},
  journal = {Phys. Rev. Lett.},
  volume = {106},
  issue = {5},
  pages = {054102},
  numpages = {4},
  year = {2011},
  month = {Feb},
  publisher = {American Physical Society},
  doi = {10.1103/PhysRevLett.106.054102},
  url = {https://link.aps.org/doi/10.1103/PhysRevLett.106.054102}
}

@article{nisoli2021color,
  title={The color of magnetic monopole noise},
  author={Nisoli, Cristiano},
  journal={Europhysics Letters},
  volume={135},
  number={5},
  pages={57002},
  year={2021},
  publisher={EDP Sciences, IOP Publishing and Societ{\`a} Italiana di Fisica},
  url = {https://iopscience.iop.org/article/10.1209/0295-5075/ac2654/meta?casa_token=nRWs1HxV7J8AAAAA:8GpaTm9bBE9aMYBMa4VMRVRkU6bLdWZrT_bAAv31yFEJaVFztauswoinCJqYTICS0UaC6MV6SWgGZ-BINeBDxWbo2jo}
}

@article{raumonop,
  title = {Thermal conductivity of square ice},
  author = {Sutcliffe, Ruairidh and Rau, Jeffrey G.},
  journal = {Phys. Rev. B},
  volume = {105},
  issue = {10},
  pages = {104405},
  numpages = {19},
  year = {2022},
  month = {Mar},
  publisher = {American Physical Society},
  doi = {10.1103/PhysRevB.105.104405},
  url = {https://link.aps.org/doi/10.1103/PhysRevB.105.104405}
}

@article{wegner1971duality,
  title={Duality in generalized Ising models and phase transitions without local order parameters},
  author={Wegner, Franz J},
  journal={Journal of Mathematical Physics},
  volume={12},
  number={10},
  pages={2259--2272},
  year={1971},
  publisher={American Institute of Physics},
url = {https://pubs.aip.org/aip/jmp/article-abstract/12/10/2259/465334/Duality-in-Generalized-Ising-Models-and-Phase}
}

@article{Kogut-suss,
  title = {Hamiltonian formulation of Wilson's lattice gauge theories},
  author = {Kogut, John and Susskind, Leonard},
  journal = {Phys. Rev. D},
  volume = {11},
  issue = {2},
  pages = {395--408},
  numpages = {0},
  year = {1975},
  month = {Jan},
  publisher = {American Physical Society},
  doi = {10.1103/PhysRevD.11.395},
  url = {https://link.aps.org/doi/10.1103/PhysRevD.11.395}
}

@article{Kogutrev,
  title = {An introduction to lattice gauge theory and spin systems},
  author = {Kogut, John B.},
  journal = {Rev. Mod. Phys.},
  volume = {51},
  issue = {4},
  pages = {659--713},
  numpages = {0},
  year = {1979},
  month = {Oct},
  publisher = {American Physical Society},
  doi = {10.1103/RevModPhys.51.659},
  url = {https://link.aps.org/doi/10.1103/RevModPhys.51.659}
}

@article{ChakrDFL1,
  title = {Disorder-free localization transition in a two-dimensional lattice gauge theory},
  author = {Chakraborty, Nilotpal and Heyl, Markus and Karpov, Petr and Moessner, Roderich},
  journal = {Phys. Rev. B},
  volume = {106},
  issue = {6},
  pages = {L060308},
  numpages = {5},
  year = {2022},
  month = {Aug},
  publisher = {American Physical Society},
  doi = {10.1103/PhysRevB.106.L060308},
  url = {https://link.aps.org/doi/10.1103/PhysRevB.106.L060308}
}

@article{Vitelliglass,
  title = {Nonreciprocal Spin-Glass Transition and Aging},
  author = {Garcia Lorenzana, Giulia and Altieri, Ada and Biroli, Giulio and Fruchart, Michel and Vitelli, Vincenzo},
  journal = {Phys. Rev. Lett.},
  volume = {135},
  issue = {18},
  pages = {187402},
  numpages = {10},
  year = {2025},
  month = {Oct},
  publisher = {American Physical Society},
  doi = {10.1103/PhysRevLett.135.187402},
  url = {https://link.aps.org/doi/10.1103/PhysRevLett.135.187402}
}

@article{ChakrDFL2,
  title = {Spectral Response of Disorder-Free Localized Lattice Gauge Theories},
  author = {Chakraborty, Nilotpal and Heyl, Markus and Karpov, Petr and Moessner, Roderich},
  journal = {Phys. Rev. Lett.},
  volume = {131},
  issue = {22},
  pages = {220402},
  numpages = {6},
  year = {2023},
  month = {Nov},
  publisher = {American Physical Society},
  doi = {10.1103/PhysRevLett.131.220402},
  url = {https://link.aps.org/doi/10.1103/PhysRevLett.131.220402}
}

@article{osborne2024quantum,
  title={Quantum Many-Body Scarring in $2+ 1$ D Gauge Theories with Dynamical Matter},
  author={Osborne, Jesse and McCulloch, Ian P and Halimeh, Jad C},
  journal={arXiv:2403.08858},
  year={2024},
url = {https://arxiv.org/abs/2403.08858}
}

@article{gyawali2410observation,
  title={Observation of disorder-free localization using a (2+ 1) D lattice gauge theory on a quantum processor},
  author={Gyawali, Gaurav and Kumar, Shashwat and Lensky, Yuri D and Rosenberg, E and Szasz, A and Cochran, T and Chen, R and Karamlou, AH and Kechedzhi, K and Berndtsson, J and others},
  journal={arXiv:2410.06557},
    url = {https://arxiv.org/abs/2410.06557}
}

@article{homeier2025prethermal,
  title={Prethermal gauge structure and surface growth in $\mathbb{Z}_2$ lattice gauge theories},
  author={Homeier, Lukas and Pizzi, Andrea and Zhao, Hongzheng and Halimeh, Jad C and Grusdt, Fabian and Rey, Ana Maria},
  journal={arXiv:2510.12800},
  year={2025},
url = {https://arxiv.org/abs/2510.12800}
}

@article{chakraborty2025fractional,
  title={Fractional diffusion without disorder in two dimensions},
  author={Chakraborty, Nilotpal and Heyl, Markus and Moessner, Roderich},
  journal={arXiv:2504.00074},
  year={2025},
url = {https://arxiv.org/abs/2504.00074}
}

@article{hallen2022dynamical,
  title={Dynamical fractal and anomalous noise in a clean magnetic crystal},
  author={Hall{\'e}n, Jonathan N and Grigera, Santiago A and Tennant, D Alan and Castelnovo, Claudio and Moessner, Roderich},
  journal={Science},
  volume={378},
  number={6625},
  pages={1218--1221},
  year={2022},
  publisher={American Association for the Advancement of Science},
  url = {https://www.science.org/doi/abs/10.1126/science.add1644}
}

@article{TCMglass,
  title = {Structural magnetic glassiness in the spin ice ${\mathrm{Dy}}_{2}{\mathrm{Ti}}_{2}{\mathrm{O}}_{7}$},
  author = {Samarakoon, Anjana M. and Sokolowski, Andr\'e and Klemke, Bastian and Feyerherm, Ralf and Meissner, Michael and Borzi, R. A. and Ye, Feng and Zhang, Qiang and Dun, Zhiling and Zhou, Haidong and Egami, T. and Hall\'en, Jonathan N. and Jaubert, Ludovic and Castelnovo, Claudio and Moessner, Roderich and Grigera, S. A. and Tennant, D. Alan},
  journal = {Phys. Rev. Res.},
  volume = {4},
  issue = {3},
  pages = {033159},
  numpages = {8},
  year = {2022},
  month = {Aug},
  publisher = {American Physical Society},
  doi = {10.1103/PhysRevResearch.4.033159},
  url = {https://link.aps.org/doi/10.1103/PhysRevResearch.4.033159}
}

@article{castelnovoreview,
  title={Spin ice, fractionalization, and topological order},
  author={Castelnovo, C and Moessner, R and Sondhi, Shivaji Lal},
  journal={Annu. Rev. Condens. Matter Phys.},
  volume={3},
  number={1},
  pages={35--55},
  year={2012},
  publisher={Annual Reviews},
  url = {https://www.annualreviews.org/content/journals/10.1146/annurev-conmatphys-020911-125058}
}

@article{Banscar,
  title = {Sublattice scars and beyond in two-dimensional $U(1)$ quantum link lattice gauge theories},
  author = {Sau, Indrajit and Stornati, Paolo and Banerjee, Debasish and Sen, Arnab},
  journal = {Phys. Rev. D},
  volume = {109},
  issue = {3},
  pages = {034519},
  numpages = {17},
  year = {2024},
  month = {Feb},
  publisher = {American Physical Society},
  doi = {10.1103/PhysRevD.109.034519},
  url = {https://link.aps.org/doi/10.1103/PhysRevD.109.034519}
}

@article{halimeh2025quantum,
  title={Quantum simulation of out-of-equilibrium dynamics in gauge theories},
  author={Halimeh, Jad C and Mueller, Niklas and Knolle, Johannes and Papi{\'c}, Zlatko and Davoudi, Zohreh},
  journal={arXiv preprint arXiv:2509.03586},
  year={2025},
url = {https://arxiv.org/abs/2509.03586}
}

@article{KarpovDFL,
  title = {Disorder-Free Localization in an Interacting 2D Lattice Gauge Theory},
  author = {Karpov, P. and Verdel, R. and Huang, Y.-P. and Schmitt, M. and Heyl, M.},
  journal = {Phys. Rev. Lett.},
  volume = {126},
  issue = {13},
  pages = {130401},
  numpages = {6},
  year = {2021},
  month = {Apr},
  publisher = {American Physical Society},
  doi = {10.1103/PhysRevLett.126.130401},
  url = {https://link.aps.org/doi/10.1103/PhysRevLett.126.130401}
}

@article{ben2025many,
  title={Many-body cages: disorder-free glassiness from flat bands in Fock space, and many-body Rabi oscillations},
  author={Ben-Ami, Tom and Heyl, Markus and Moessner, Roderich},
  journal={arXiv:2504.13086},
  year={2025},
url = {https://arxiv.org/abs/2504.13086}
}

@article{SmithDFL,
  title = {Dynamical localization in ${\ensuremath{\mathbb{Z}}}_{2}$ lattice gauge theories},
  author = {Smith, Adam and Knolle, Johannes and Moessner, Roderich and Kovrizhin, Dmitry L.},
  journal = {Phys. Rev. B},
  volume = {97},
  issue = {24},
  pages = {245137},
  numpages = {18},
  year = {2018},
  month = {Jun},
  publisher = {American Physical Society},
  doi = {10.1103/PhysRevB.97.245137},
  url = {https://link.aps.org/doi/10.1103/PhysRevB.97.245137}
}

@article{sirote2024emergent,
  title={Emergent disorder and mechanical memory in periodic metamaterials},
  author={Sirote-Katz, Chaviva and Shohat, Dor and Merrigan, Carl and Lahini, Yoav and Nisoli, Cristiano and Shokef, Yair},
  journal={Nature Communications},
  volume={15},
  number={1},
  pages={4008},
  year={2024},
  publisher={Nature Publishing Group UK London},
url = {https://www.nature.com/articles/s41467-024-47780-w}
}

@article{coulais2016combinatorial,
  title={Combinatorial design of textured mechanical metamaterials},
  author={Coulais, Corentin and Teomy, Eial and De Reus, Koen and Shokef, Yair and Van Hecke, Martin},
  journal={Nature},
  volume={535},
  number={7613},
  pages={529--532},
  year={2016},
  publisher={Nature Publishing Group UK London},
url = {https://www.nature.com/articles/nature18960
}
}

@article{meeussen2020topological,
  title={Topological defects produce exotic mechanics in complex metamaterials},
  author={Meeussen, Anne S and O{\u{g}}uz, Erdal C and Shokef, Yair and Hecke, Martin van},
  journal={Nature Physics},
  volume={16},
  number={3},
  pages={307--311},
  year={2020},
  publisher={Nature Publishing Group UK London},
url = {https://www.nature.com/articles/s41567-019-0763-6}
}

@article{merrigantop,
  title = {Topologically protected steady cycles in an icelike mechanical metamaterial},
  author = {Merrigan, Carl and Nisoli, Cristiano and Shokef, Yair},
  journal = {Phys. Rev. Res.},
  volume = {3},
  issue = {2},
  pages = {023174},
  numpages = {14},
  year = {2021},
  month = {Jun},
  publisher = {American Physical Society},
  doi = {10.1103/PhysRevResearch.3.023174},
  url = {https://link.aps.org/doi/10.1103/PhysRevResearch.3.023174}
}

@article{Jorgeflows,
  title = {Active-Hydraulic Flows Solve the Six-Vertex Model (and Vice Versa)},
  author = {Jorge, Camille and Bartolo, Denis},
  journal = {Phys. Rev. Lett.},
  volume = {134},
  issue = {18},
  pages = {188302},
  numpages = {6},
  year = {2025},
  month = {May},
  publisher = {American Physical Society},
  doi = {10.1103/PhysRevLett.134.188302},
  url = {https://link.aps.org/doi/10.1103/PhysRevLett.134.188302}
}

@article{jorge2024active,
  title={Active hydraulics laws from frustration principles},
  author={Jorge, Camille and Chardac, Am{\'e}lie and Poncet, Alexis and Bartolo, Denis},
  journal={Nature Physics},
  volume={20},
  number={2},
  pages={303--309},
  year={2024},
  publisher={Nature Publishing Group UK London},
url = {https://www.nature.com/articles/s41567-023-02301-2}
}

@book{moessner2021topological,
  title={Topological phases of matter},
  author={Moessner, Roderich and Moore, Joel E},
  year={2021},
  publisher={Cambridge University Press}
}

@article{antal2002repeated,
  title={Repeated bond traversal probabilities for the simple random walk},
  author={Antal, T and Hilhorst, HJ and Zia, RKP},
  journal={Journal of Physics A: Mathematical and General},
  volume={35},
  number={39},
  pages={8145},
  year={2002},
  publisher={IOP Publishing},
url = {https://iopscience.iop.org/article/10.1088/0305-4470/35/39/301/meta?casa_token=SVk9NVyCXl0AAAAA:XMQQAKe3HpkDJCL2y46B_0W2M6Fhlzt3CgDmHmBKjhKUmL1KelR13TEScPpcpLdwnytqhfDEZHZOO0PNybTpPROukmG7}
}

@book{udagawa2021spin,
  title={Spin Ice},
  author={Udagawa, Masafumi and Jaubert, Ludovic and others},
  volume={197},
  year={2021},
  publisher={Springer}
}

@misc{supp,
  note = "See Supplemental Material at
    URL-will-be-inserted-by-publisher."
}

\end{document}